\documentclass[pra,reprint,amssymb]{revtex4-1}%

\usepackage{amsmath,amssymb}
\usepackage{revsymb4-1}
\usepackage[cmyk]{xcolor}
\usepackage[%
	colorlinks=true,%
	linkcolor=blue,%
	citecolor=blue,%
	urlcolor=blue
	]{hyperref}%
\usepackage{graphicx}%
\usepackage{here}
\usepackage{dcolumn}%
\usepackage{bm}%
\usepackage{revsymb}
\usepackage{braket}
\usepackage{layouts} 
\usepackage{color}
 
\def\equationautorefname~#1\null{\textrm{~(#1)\;}\null}
\def\figureautorefname~#1\null{Fig.~#1\null}

\begin{document}

\title{Decay of phase-imprinted dark soliton in Bose--Einstein condensate \\ at non-zero temperature}
\author{Hiroki Ohya, Shohei Watabe, and Tetsuro Nikuni}
\begin{abstract}
      We study relaxation dynamics of dark soliton, created by a phase-imprinted method, in a two-dimensional trapped Bose--Einstein condensate at non-zero temperatures by using the projected Gross-Pitaevskii equation. 
      At absolute zero temperature, a dark soliton is known to decay with a snake instability. At non-zero temperature, as we expected, we find that this snake instability cannot be clearly seen as in the absolute zero temperature case because of the presence of thermal fluctuations. However, we find that the decay rate, the half width of the overlap integral with respect to the phase-imprinted initial state, shows a power low decay as a function of the energy and finally remains a non-zero value. 
      \end{abstract}

\affiliation{Tokyo University of Science, 1-3 Kagurazaka, Shinjuku-ku, Tokyo, 162-9601, Japan}
\maketitle



A Bose--Einstein condensate (BEC)---one of the macroscopic quantum phenomena---exhibits a matter wave nature. 
A soliton in the BEC is a simple but attractive structure of the matter wave, and that is created in ultracold atoms by using the fascinating method which is called phase-imprinting~\cite{Burger1999,Denschlag2000,Becker2008}. 
The soliton in a BEC is stable in a (quasi-)one-dimensional regime, where the transverse modulation instability is suppressed~\cite{Brand2002}. 
The transverse confinement is the weaker, what amounts to increasing extra dimensionality, the more unstable the soliton is, which is named the snake instability after the transverse fluctuation of the soliton~\cite{Mamaev1996,Feder2000,Dutton2001}. 

Studying the soliton in a BEC at absolute zero temperature is very interesting, since it can be well written by the nonlinear Schr\"odinger equation called the Gross--Pitaevskii (GP) equation~\cite{PitaevskiiBOOK}. 
Non-zero temperature effect of the soliton in a BEC is also a quite important issue to understand behavior of the matter wave if we are considering the case of real experimental situations. 
This non-zero temperature effect is numerically studied in a highly elongated trap case, where the soliton is relatively stable~\cite{Jackson2007}. 
In that study, where the Zaremba-Nikuni-Griffin (ZNG) equation, coupled equations of the generalized GP equation and the Boltzmann equation for the thermal cloud, is used, it was found that with temperature, the oscillation amplitude of a soliton increases and the soliton depth and energy decrease~\cite{Jackson2007}. 
However, higher dimensional effect is open on a phase-imprinted soliton in a BEC at non-zero temperature, where a soliton shows the snake instability at $T = 0$. 

In this paper, we study relaxation dynamics of a phase-imprinted dark soliton in a two-dimensional trapped Bose--Einstein condensate at non-zero temperature after a sudden quench of the phase-imprinting. 
Since two-dimensional system has stronger thermal fluctuations, we employ the projected Gross-Pitaevskii equation (PGPE)~\cite{Davis2001,Davis2002,Davis2003,Blakie2005,Blakie2007,Blakie2008AiP,Blakie2008PRE}, instead of the mean-field-type ZNG equation. 
The depth of the soliton evaluated in Ref.~\cite{Jackson2007} may not be useful for the present study because of the strong fluctuations. 
Instead of the depth, since the BEC is a matter wave, we estimate the decay of the phase-imprinted soliton through the fidelity, the overlapping factor with the phase-imprinted initial state. 
We find that the fidelity smoothly decays and oscillates around a non-zero value, where the fidelity at $T = 0$ remains almost unity while the snake instability emerges and sharply drops when a soliton decays into vortices. 
We also find that with the energy, the half-width of the fidelity shows a power law decay and remains a constant value. 
Our numerical simulation suggests that this soliton decay rate obeys an inverse square law. 
The present study will promote further study of the soliton in matter waves at non-zero temperature.


We consider a two-dimensional weakly interacting Bose gas with an atomic mass $m$ confined in a harmonic trap at non-zero temperatures. 
The dynamics of the Bose-field $\Psi_{} ({\bf r}, t)$ in the classical region $({\rm C})$ is well described by the PGPE~\cite{Davis2001,Davis2002,Davis2003,Blakie2005,Blakie2007,Blakie2008AiP,Blakie2008PRE}
\begin{align}
i \hbar \frac{\partial \Psi_{} ({\bf r} ,t)}{\partial t} = \hat H_{0} ({\bf r})\Psi ({\bf r} ,t) + {\mathcal P} [U |\Psi ({\bf r}, t)|^{2} \Psi ({\bf r}, t) ], 
\label{PGPE}
\end{align} 
where 
\begin{align}
\hat H_{0} ({\bf r}) = - \frac{\hbar^{2}}{2m} \nabla^{2} + V_{\rm trap} ({\bf r}), 
\end{align}
with the harmonic trap potential of frequencies $\omega_{x,y}$, given by 
\begin{align}
 V_{\rm trap} ({\bf r}) =  \frac{m\omega_{x}^{2}}{2} x^{2} + \frac{m\omega_{y}^{2}}{2} y^{2}. 
\end{align}
The term ${\mathcal P} [U |\Psi ({\bf r}, t)|^{2} \Psi ({\bf r}, t) ]$ with the contact interaction strength $U$ provides the non-linear term projected to the classical region, where the projection operator applied to a function $F ({\bf r} , t)$ is given by 
\begin{align}
{\mathcal P} [F ({\bf r}, t) ] \equiv \sum\limits_{{\bf n}\in {\rm C}} \phi_{\bf n} ({\bf r}) \int d {\bf r}' \phi_{\bf n}^{*} ({\bf r}') F ({\bf r}', t). 
\end{align} 
The function $\phi_{\bf n} ({\bf r})$ is an a single-particle eigenstate of $\hat H_{0}$ with the quantum numbers ${\bf n}$. The summation is restricted to the eigenmodes in the classical region that we take as the region where the mean occupation number of the single-particle eigenstates is more than $n_{\rm cut }$. 
The time-dependence of the classical Bose-field is determined by a coefficient $c_{\bf n} (t)$ through $\Psi ({\bf r}, t) = \sum\limits_{{\bf n} \in C} c_{\bf n} (t) \phi_{\bf n}^{} ({\bf r})$ with ${\bf n} \in {\rm C}$.

\begin{figure}[tbp] 
\begin{center}
\includegraphics[width=8cm]{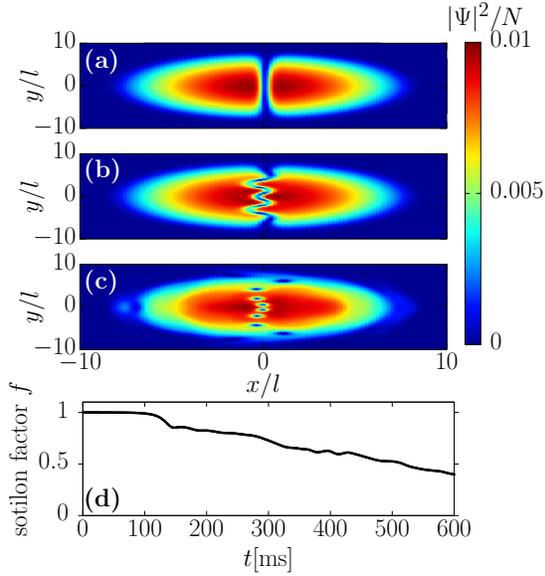}
\end{center} 
\caption{
Decay of a soliton in the absolute zero temperature case $T=0$, simulated by the Gross-Pitaevskii equation. 
(a) $t = 0$ms. (b) $t = 82$ms. (c) $t = 150$ms. (d) The time-dependence of the fidelity $f$. 
We used $\omega_{x,y} = 2\pi \times 100$Hz, $N = 1.0 \times 10^{3}$ $^{85}$Rb atoms, and $U = 3.5\hbar^{2} / m$. 
 }
\label{fig1}
\end{figure}

We first prepare a pre-initial state and determine the classical region following the semiclassical Hartree--Fock approximation and the Thomas--Fermi approximation~\cite{Blakie2007,Sato2009}. 
The pre-initial coefficient in the classical regime $c_{\bf n}^{\rm pre}$ at the non-zero temperature $T$, the state with which is to relax to an equilibrium state through the PGPE, is determined by $c_{n}^{\rm pre} = \exp(i \theta_{\rm rand})\sqrt{n_{\rm B} (\epsilon_{\bf n}, T, \mu_{\rm TF})}$,  
where $\theta_{\rm rand}$ is the randomized phase factor, and $n_{\rm B} (\epsilon_{{\bf n}}, T, \mu_{\rm TF}) = 1/\{ \exp[ (\epsilon_{\bf n} - \mu_{\rm TF})/(k_{\rm B}T) ] - 1\}$ is the Bose-distribution function at the single-particle eigenenergy $\epsilon_{{\bf n}} = (n_{x} + 1/2) \hbar \omega_{x} + (n_{y} + 1/2) \hbar \omega_{y}$ with $n_{x,y} = 0, 1, 2, \cdots$. The chemical potential $\mu_{\rm TF}$ in the Thomas--Fermi approximation is determined by 
\begin{align}
\mu_{\rm TF} = \sqrt{\frac{m\omega_{x} \omega_{y} U}{\pi} (N_{\rm tot} - N')}, 
\end{align}
where $N'$ is the number of the noncondensate particle 
\begin{align}
N' = \int d \epsilon \rho (\epsilon ) n_{\rm B} ( \epsilon, T, \mu_{\rm TF}), 
\end{align}
with the semiclassical density of states in the Hartree--Fock approximation 
\begin{align}
\rho (\epsilon) = \int \frac{d{\bf r} d {\bf p}}{(2\pi\hbar)^{2}} 
\delta \left ( \epsilon - \left [ \frac{{\bf p}^{2}}{2m} + V_{\rm trap} ({\bf r}) + 2 U n_{0} ({\bf r})  \right ] \right ). 
\end{align} 
The condensate density $n_{0} ({\bf r})$ in the Thomas--Fermi approximation is given by 
\begin{align}
n_{0} ({\bf r}) = 
\left \{ 
\begin{array}{ll}
[\mu_{\rm TF} - V_{\rm trap} ({\bf r})] / U,  & \mu_{\rm TF} - V_{\rm trap} ({\bf r}) \geq 0, 
\\ 
0,  & \mu_{\rm TF} - V_{\rm trap} ({\bf r}) < 0. 
\end{array}
\right .  
\end{align} 
We self-consistently solve those equations to determine $\mu_{\rm HF}$ at a given temperature $T$. 
The energy cutoff $\epsilon_{\rm cut}$, which the classical regime is below, is determined by the condition $n_{\rm cut} = n_{\rm B} (\epsilon_{\rm cut}, T, \mu_{\rm TF})$, where 
we take $n_{\rm cut } = 3$~\cite{Blakie2007,Sato2009}.   
Given $c_{\bf n}^{\rm pre}$ as an initial sate, we solve the PGPE for sufficiently long time steps, which may provide an equilibrium state, where we name this time $t = 0^{-}$. 

\begin{figure}[tbp] 
\begin{center}
\includegraphics[width=8cm]{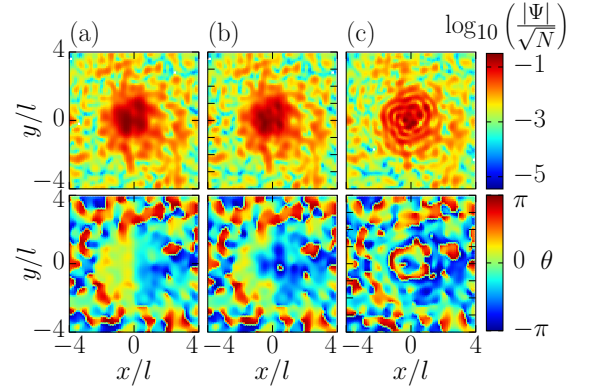}
\end{center} 
\caption{
Decay of a phase-imprinted soliton simulated by the Projected Gross-Pitaevskii equation at $E = 176\hbar \omega$. 
(a) $t = 0$ms. (b) $ = 30$ms. (c) $t = 120$ms. 
In this simulation, we used $\omega_{x,y} = 2\pi \times 100$Hz, $N = 1.0 \times 10^{4}$ $^{85}$Rb atoms, and $U = 1.0\hbar^{2} / m$. 
 }
\label{fig2}
\end{figure}

The soliton is created by the phase-imprinting method~\cite{Denschlag2000}, where the phase is imprinted by a spatially varying light-shift potential which atoms experience. 
In this paper, we investigate relaxation dynamics of a phase-imprinted dark soliton in a non-zero temperature BEC after a sudden quench of the phase-imprinting. 
We thus prepare the phase-printed initial state $\Psi ({\bf r},t = 0^{+})$ from the state $\Psi ({\bf r}, 0^{-})$,  through the relation 
\begin{align}
\Psi ({\bf r}, t =0^{+}) = 
\begin{cases}
\Psi ({\bf r}, t =0^{-}), & x < 0, 
\\ 
\exp (i\theta) \Psi ({\bf r}, t =0^{-}), & x \geq 0, 
\end{cases}
\end{align} 
where the notch of the soliton is positioned along the $y$-axis. 
By solving the PGPE starting from the phase-imprinted initial state $\Psi ({\bf r},t = 0^{+})$, we study the decay of a phase-imprinted soliton at non-zero temperatures. 
In this paper, we study the phase-imprinted dark-soliton case $\theta = \pi$.

We here summarize the method we applied in this paper to prepare the phase-imprinted initial soliton state $\Psi ({\bf r}, t = 0^{+})$ from an equilibrium state $\Psi ({\bf r}, t = 0^{-})$. 
It is convenient to introduce the normalized formula, such as $\tilde x \equiv x / l$, $\tilde y \equiv y /l$ and $\tilde \Psi \equiv \Psi l/ \sqrt{N}$, with $l \equiv \sqrt{\hbar /(m \omega)}$, where in the following we omit the tilde for the simplicity. 
The initial states can be written as 
\begin{align}
\Psi^{\pm} ({\bf r}) \equiv \Psi (x, y, t = 0^{\pm}) = \sum\limits_{n_{x,y} \in C} c_{n_{x}, n_{y}}^{\pm} \phi_{n_x} (x)  \phi_{n_y} (y), 
\end{align}
where $\phi_{n} (x)$ is the single-particle eigenstate for a harmonic trap in the dimensionless form, given by 
\begin{align}
\phi_{n} (x) = \left ( \frac{1}{\pi} \right )^{1/4}\frac{1}{\sqrt{2^{n} n!}}H_{n}(x) \exp \left ( {-\frac{x^{2}}{2}} \right ). 
\end{align}
Here, $H_{n}(x)$ is the Hermite polynomials. 

We determine the coefficient $c_{n_{x}, n_{y}}^{+}$ from $c_{n_{x}, n_{y}}^{-}$ by using the phase-imprinting condition 
\begin{align}
\Psi^{+} ({\bf r}) = \begin{cases}
\Psi^{-} ({\bf r}),  & x \geq 0,  
\\
e^{i\theta} \Psi^{-} ({\bf r}), & x<0. 
\end{cases}
\end{align}
Indeed, we have a relation 
\begin{align}
c_{n_{x}, n_{y}}^{+} = & 
\int_{-\infty}^{\infty} dx  \int_{-\infty}^{\infty}  dy \phi^{*}_{n_{x}} (x) \phi^{*}_{n_{y} }(y) \Psi^{+} ({\bf r}) 
\\
\equiv & C_{n_{x}, n_{y}}^{(+)} + \exp (i \theta) C_{n_{x}, n_{y}}^{(-)}, 
\end{align}
where 
\begin{align}
C_{n_{x}, n_{y}}^{(\pm)} \equiv & \pm \int_{0}^{\pm \infty} dx  \int_{-\infty}^{\infty}  dy \phi^{*}_{n_{x}} (x) \phi^{*}_{n_{y} }(y) \Psi^{-} ({\bf r}) . 
\end{align}

By using the orthonormal condition with respect to $\phi_{n_{y}} (y)$, 
we have 
\begin{align}
C_{n_{x}, n_{y}}^{(\pm)} \equiv & \sum\limits_{n_{x}' \in C}  \frac{1}{\sqrt{\pi 2^{(n_{x} + n_{x}')} n_{x}! n_{x}'!} } I_{n_{x}, n_{x}'}^{(\pm)} c_{n_{x}',n_{y}}^{-}, 
\end{align}
where 
\begin{align}
I_{n_{x}, n_{x}'}^{(\pm)} \equiv \pm \int_{0}^{\pm \infty} dx  H_{n_{x} }(x) H_{n_{x}'} (x) \exp ( - x^{2}). 
\end{align}
By using the Rodrigues's formula of the Hermite polynomials, we can determine the term $I_{n_{x}, n_{x}'}^{(\pm)}$ through the recurrence relation 
\begin{align}
I_{n_{x}, n_{x}'}^{(\pm)} = \mp f_{n_{x}, n_{x}'} + 2 n_{x} I_{n_{x}-1, n_{x}'-1}^{(\pm)}. 
\end{align}
Only if the condition $(n_{x}, n_{x}') = ({\rm even}, {\rm odd})$ holds, the term $f_{n_{x}, n_{x}'}$ is then given by 
\begin{align}
f_{n_{x}, n_{x}'} 
= & 
- H_{n_{x}} (x=0) H_{n_{x}'-1} (x=0) 
\\ 
= & (-1)^{ \frac{n_{x}'-1}{2} } \left ( \frac{n_{x}}{2} + 1 \right )_{n_{x}/2} \left ( \frac{n_{x}'+1}{2} \right )_{(n_{x}'-1)/2}, 
\end{align}
where the notation $(a)_{b}$ represents the Pochhammer symbol, and otherwise we have $f_{n_{x}, n_{x}'} = 0$.
The first terms in the recurrence relation, $I_{n_{x}}^{(\pm)} \equiv I_{n_{x},0}^{(\pm)} = I_{0,n_{x}}^{(\pm)}$, are given by 
\begin{align}
I_{n_{x}}^{(\pm)} 
= 
\begin{cases}
\sqrt{\pi}/2, & n_{x} = 0,
\\ 
\pm 1, & n_{x} = 1,
\\ 
0, & n_{x} = 2, 4, 6, \cdots,
\\ 
2 (2 - n_{x}) I_{n_{x}-2}^{(\pm)},  & n_{x} = 3, 5, 7, \cdots . 
\end{cases}
\end{align}

 
Before showing non-zero temperature results, we summarize characteristics of soliton decay in the two-dimensional BEC in the absolute zero temperature case. 
Figures~\ref{fig1} (a), (b), (c) shows the density plot of the condensate density, showing a typical example of soliton decay, which is obtained by solving the GP equation 
\begin{align}
i \hbar \frac{\partial \Psi ({\bf r} ,t)}{\partial t} = \hat H_{0} ({\bf r})\Psi ({\bf r} ,t) + U |\Psi ({\bf r}, t)|^{2} \Psi ({\bf r}, t). 
\label{PGPE}
\end{align} 
Although a soliton is stable in the (quasi-)one-dimensional system, extra dimensionality imposed to the system stimulates instability of the soliton. 
In particular, the dark soliton is known to decay into vortices through the snake instability~\cite{Mamaev1996,Feder2000,Dutton2001,Brand2002}. (Panels (a), (b), (c) in Fig.~\ref{fig1}.) 

\begin{figure}[tbp] 
\begin{center}
\includegraphics[width=8cm]{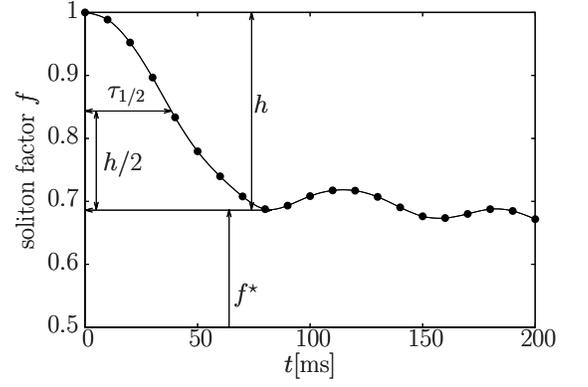}
\end{center} 
\caption{Time-dependence of the fidelity $f$ at $E = 176\hbar \omega$. 
The half width $\tau_{1/2}$ is defined by $f (t = \tau_{1/2}) \equiv 1-h/2$, 
where the height $h$ is given by $h \equiv 1 - f^{\star}$ with the fidelity $f^{*}$ that takes the first minimal. 
The parameters $\omega_{x,y}$, $N$, and $U$ are the same as in Fig.~\ref{fig2}. 
 }
\label{fig3}
\end{figure}

Instability of a soliton is to be qualitatively understood through the density profile~\cite{Mamaev1996,Feder2000,Dutton2001}, and to be quantitatively evaluated through a soliton depth~\cite{Jackson2007}. 
On the other hand, since the classical Bose-field, which is to be treated in the PGPE, involves complex fluctuations, we here extract character of the soliton decay through the fidelity, the overlap integral between the initially prepared soliton state and its time-evolved state~\cite{JSato2012}
\begin{align}
f (t > 0) \equiv 
\left | 
\int d {\bf r} \Psi^{*} ({\bf r} ,t > 0) \Psi^{} ({\bf r}, t = 0^{+}) 
\right |^{2} . 
\end{align}
When the snake instability appears, the overlap sill maintain a quite high value $f \simeq 0.996$ at $t = 82$ms (panels (b) and (d) in Fig~\ref{fig1}).
However, when the soliton decays into vortices after undergoing the snake instability (at $t = 150$ms in panels (c) in Fig~\ref{fig1}), 
the fidelity sharply drops, where a macroscopic quantum tunneling emerges and a plateau of the fidelity in initial time steps may be regarded as its tunneling time. 
The fidelity then decreases almost monotonically, which indicates the initial soliton state is disappearing after decay of the soliton into vortices. 

We here show the non-zero temperature case obtained by solving the PGPE (Fig.~\ref{fig2}). 
In this non-zero temperature case, the snake nature cannot be clearly seen in notch of the density profile. 
On the other hand, the phase contrast moves at the high density regime (panel (b) in Fig.~\ref{fig2} at $t = 30$ms). 
Finally, the soliton decay into many defects (panel (c) in Fig.~\ref{fig2} at $t = 120$ms). 
Although it is difficult to extract the decay character from the density and phase profiles because of the presence of fluctuations, 
we then evaluate character of soliton decay from the fidelity $f$. 
In contrast to the absolute zero temperature case where the plateau can be seen in the initial time steps, 
the fidelity $f$ in the PGPE smoothly decays starting from $f =1$ without a plateau, and oscillates around a certain value (Fig.~\ref{fig3}). 

\begin{figure}[tbp] 
\begin{center}
\includegraphics[width=8cm]{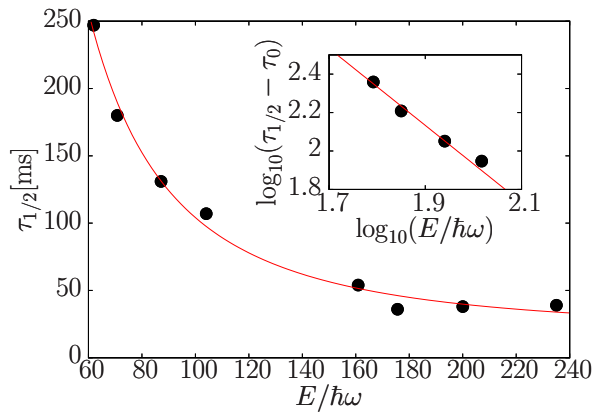}
\end{center} 
\caption{Energy-dependence of the half-width $\tau_{1/2}$ of the fidelity $f$. 
The line is a guide to an eye, a fitting function $\tau_{1/2} (E) = \tau_{0} + a [E/(\hbar \omega) ]^{-b}$ with fitting parameters $\tau_{0} \simeq 18$ms, $a \simeq 870$s, and $b \simeq 2.0$. 
The parameters $\omega_{x,y}$, $N$, and $U$ are the same as in Fig.~\ref{fig2}. 
 }
\label{fig4}
\end{figure}

We extract decay rate of the soliton from the half width $\tau_{1/2}$ defined by $f (t = \tau_{1/2}) \equiv 1-h/2$. 
Here, the height $h$ is given by $h \equiv 1 - f^{\star}$, where $f^{*}$ is the fidelity that takes the first minimal (Fig.~\ref{fig3}). 
The energy dependence of the half width $\tau_{1/2}$ is shown in Fig.~\ref{fig4}, 
where the energy $E$ is here estimated by using the phase-imprinted soliton state $\Psi ({\bf r}, \tau = 0^{+})$ through the form
\begin{align}
E = \int d{\bf r} \left [ \Psi_{}^{*} ({\bf r}) \hat H_{0} ({\bf r})\Psi_{} ({\bf r} ) + \frac{U}{2} |\Psi_{} ({\bf r})|^{4} \right ]. 
\label{PGPEE}
\end{align} 
The half-width $\tau_{1/2}$ shows a power law decay with respect to $E$ and finally remains a non-zero value, which is well fitted by a function $\tau_{1/2} (E) = \tau_{0} + a [E/(\hbar \omega) ]^{-b}$ with fitting parameters $\tau_{0} \simeq 18$ms, $a \simeq 870$s, and $b \simeq 2.0$. 
This numerical result suggests that the soliton decay rate may show the inverse square law with the energy $E$.

The phase-imprinted soliton state is unstable, and its classical Bose-field $\Psi ({\bf r}, t=0^{+})$ relaxes to a new state. 
In the higher-energy regime, the classical Bose-field $\Psi ({\bf r}, t = 0^{+})$ in the initial phase-imprinted states is constructed with more modes, which largely overlaps with the high-energy relaxed state. It may provide the energy-independent decay rate of the phase-imprinted state at higher energies.

In conclusions, we have studied relaxation dynamics of the phase-imprinted dark soliton in a two-dimensional trapped Bose--Einstein condensate (BEC) at non-zero temperatures by numerically solving the projected Gross-Pitaevskii equation. 
In absolute zero temperature case, a soliton decays into vortices through the snake instability, which can be clearly seen in the density profile. 
In contrast to this case, the density profile is not useful to observe the decay of the soliton in the two-dimensional case at non-zero temperatures, because of strong thermal fluctuations. 
Since the BEC is well described by the matter wave nature, we evaluated the soliton decay through the fidelity, the overlapping integral with the phase-imprinted initial soliton state. 
We found that the soliton decay rate, the half-width of the fidelity, shows a power law decay and remains a constant value as a function of the energy that is estimated by the initial phase-imprinted state. 
Our numerical simulation suggests that this soliton decay rate obeys an inverse square law. 
An interesting open issue related to this study is to uncover effect of the two-dimensional Berezinskii-Kosterlitz-Thouless transition with vortex pairs~\cite{Hadzibabic2006,Hung2011,Fletcher2015} on the relaxation dynamic of the soliton. 

\begin{acknowledgments}
Authors thanks to T. Sato for discussing implementation of the PGPE. S.W. has been supported by JSPS KAKENHI Grant No. JP16K17774. T.N. has been supported by JSPS KAKENHI Grant No. JP16K05504.
\end{acknowledgments}

\bibliographystyle{apsrev4-1}
\bibliography{library.bib}

\end{document}